\documentclass[aps,onecolumn,showpacs,amsfonts,eqsecnum,float,preprint]{revtex4} 
\usepackage{epsfig,amsmath}
\usepackage{bm}

\begin{document}

\title{Complex trajectory method in time-dependent WKB}

\author{Yair Goldfarb${}^1$, Jeremy Schiff${}^{1,2,3}$ and David J. Tannor${}^1$} \affiliation{${}^1$Department of
Chemical Physics, ${}^2$Department of Mathematics, The Weizmann
Institute of Science, Rehovot, 76100 Israel \\ \today}

\begin{abstract}
\noindent We present a significant improvement to a time-dependent
WKB (TDWKB) formulation developed by Boiron and Lombardi [JCP
{\bf108}, 3431 (1998)] in which the TDWKB equations are solved along
classical trajectories that propagate in the complex plane. Boiron
and Lombardi showed that the method gives very good agreement with
the exact quantum mechanical result as long as the wavefunction does
not exhibit interference effects such as oscillations and nodes. In
this paper we show that this limitation can be overcome by
superposing the contributions of \textit{crossing} trajectories. We
also demonstrate that the approximation improves when incorporating
higher order terms in the expansion. These improvements could make
the TDWKB formulation a competitive alternative to current
time-dependent semiclassical methods.
\bigskip

\noindent ${}^3$On sabbatical leave from Dept. of Mathematics,
Bar-Ilan University, Ramat Gan 52900, Israel.

\end{abstract}

\maketitle



\section{Introduction}
\noindent The difficulty in performing quantum mechanical
calculations of multi-dimensional systems has stimulated an
intensive and ongoing effort in the last three decades to develop
numerical tools based on semiclassical mechanics. In this context,
we refer to semiclassical mechanics as the derivation of a quantum
mechanical wavefunction or propagator via propagation of classical
(or classical-like) trajectories. From a physical point of view,
semiclassical methods try to evade the non-locality imbedded in
quantum mechanics. Mathematically speaking, semiclassical methods
aim at casting the time-dependent Schr\"odinger equation (TDSE),
which is a PDE, in terms of ODEs related to classical equations of
motion. This transformation has significant computational advantages
that can ease the inherent difficulty of multi-dimensional quantum
calculations.

The WKB method\cite{wentzel,kramers,brillouin} can be considered as
the first of the semiclassical methods. Its date of birth almost
coincides with the publication of the Schr\"odinger equation in
1926, and virtually every standard text book in quantum mechanics
has a description of the method. The basic idea of the WKB method is
to recast the wavefunction as the exponential of a function and then
replace the exponent with a power series in $\hbar$. The WKB method
is ordinarily applied to the \textit{time-independent} Schr\"odinger
equation and provides for a good approximation to the eigenstates as
long as one is not too near a classical turning point. It is only
natural that as part of the effort to develop time-dependent
semiclassical methods, a time-dependent version of the WKB method
would be explored. Surprisingly little work has been done in this
direction\cite{pauli,kurt,eu,kor,bli,rai,fis,ron,bur,spa,san,cho,bra}.
A decade ago, Boiron and Lombardi\cite{boiron} developed a complex
trajectory version of time-dependent WKB (TDWKB), which we refer to
as CTDWKB. In conventional WKB the leading order term in the phase
of the wave function is taken to be $O(\hbar^{-1})$ and the leading
order term in the amplitude is taken to be $O(\hbar^0)$. In
contrast, the CTDWKB formulation treats the amplitude and phase on
an equal footing. The price to pay for this procedure is that the
resulting classical trajectories propagate in the complex plane. The
benefits are that the results are superior to standard TDWKB and no
singularities are encountered during the integration of the equation
of motion.

The CTDWKB equations of motion can be solved analytically and yield
the \textit{exact} wavefunction for an initial Gaussian wavepacket
in a potential with up to quadratic terms. The first-order method
was tested numerically by Boiron and Lombardi for scattering of a
Gaussian wavepacket from a potential barrier. They showed that the
method produced very good results as long as the wavefunction did
not exhibit \textit{interference effects} in the form of
oscillations or nodes\cite{boiron}. In this paper we present a
simple modification to CTDWKB that provides an accurate description
of oscillations in the wavefunction. We show that complex classical
trajectories, similar to real classical trajectories, can cross in
configuration space. By superposing the contributions from two or
more crossing trajectories, interference effects are obtained. We
take CTDWKB a step further in another direction by showing that the
approximation generally improves when incorporating additional terms
in the series expansion. Since the WKB expansion is an asymptotic
series, this observation is non-trivial.

Two other semiclassical formulations that incorporate complex
trajectories should be mentioned in relation with CTDWKB. The first
is the Generalized Gaussian Wavepacket Dynamics (GGWPD) developed by
Huber, Heller and Littlejohn \cite{huber1,huber2}. One may show that
for an initial Gaussian wavepacket the equations of motion of GGWPD
are de facto identical to the equations of the first-order
approximation of CTDWKB. However the GGWPD has no generalization to
arbitrary initial wavefunctions and no systematic way to increase
the accuracy of the approximation. On the other hand, in reference
\cite{huber1}, Huber and Heller appreciate the importance of
multiple complex trajectories in obtaining interference phenomena.
Here we incorporate the idea of crossing complex trajectories into
the more general CTDWKB formulation.

The second formulation that is closely related to CTDWKB is Bohmian
Mechanics with Complex Action (BOMCA)\cite{goldfarb,goldfarb2}.
CTDWKB and BOMCA begin with the same ansatz of substituting of an
exponential function $\exp[iS/\hbar]$ into the TDSE. Similar to
CTDWKB, the BOMCA formulation incorporates equations of motion that
propagate along complex trajectories. The first-order equations of
motion of BOMCA are identical to the equations of first-order
CTDWKB. The differences between the two formulations are: (1) The
equations of motion in BOMCA are for the coefficients of spatial
derivatives of the phase. In CTDWKB the equations of motion are for
the coefficients of an $\hbar$ Taylor expansion of the phase and
their spatial derivatives. (2) Incorporating higher order terms of
the CTDWKB approximation does not effect the the results for lower
order terms since each equation of motion depends only on lower
terms of the expansion. This is not the case with BOMCA where each
equation of motion depends on both lower and higher terms resulting
in a backward feedback. (3) A result of the last difference is that
in CTDWKB the equations of motion of the trajectories remain
\textit{classical} whereas in BOMCA, the inclusion of higher orders
of the approximation affect the complex trajectories by adding a
``quantum force" that yields \textit{quantum} trajectories.

This paper is organized as follows. In section \ref{formulation} we
formulate TDWKB and the CTDWKB. Our derivation is more compact than
the Boiron-Lombardi derivation and demonstrates how to obtain the
equations of motion for higher orders of the expansion in a simple
manner. In Section \ref{num} we apply the formulation to a Gaussian
initial wavepacket propagating in a quartic double-well potential.
We demonstrate that superimposing the contributions of crossing
trajectories leads to interference effects and that incorporating
higher order terms in the expansion improves the approximation.
Section \ref{summary} is a summary and concluding remarks. Following
Boiron and Lombardi we will refer to the CTDWKB method in the body
of the paper as the complex trajectory method (CTM) for short.


\section{Formulation}
\label{formulation}

\subsection{Time-independent vs. Time-dependent WKB}

\noindent For simplicity we present the one-dimensional version of
the CTM derivation. The generalization to multi-dimensions can be
performed in a straightforward manner. The conventional WKB
derivation begins by inserting the ansatz
\begin{equation}
\label{ansatz1}
 \psi(x)=\exp\left[\frac{i}{\hbar}S(x)\right],
\end{equation}
into the \textit{time-independent} Schr\"odinger equation, where
$\hbar$ is Planck's constant divided by $2\pi$. The end result is
\begin{equation}
\label{dsdx}
\frac{1}{2m}\left(\frac{dS}{dx}\right)^{2}+V(x)-\frac{i\hbar}{2m}\frac{d^{2}S}{dx^{2}}=E,
\end{equation}
where $m$ is the mass of the particle, $V(x)$ is the potential
energy and $E$ is the eigenvalue. If we assume that $S(x)$ can be
expanded asymptotically as a polynomial in $\hbar$
\begin{equation}
\label{expansion1}
 S(x)=S_{0}(x)+\hbar S_{1}(x)+\hbar^{2}S_{2}(x)+...=\sum_{j=0}^{\infty}\hbar^{j}S_{j}(x),
\end{equation}
then, by substituting the last equation into eq.(\ref{dsdx}) and
equating powers of $\hbar$, a series of coupled ODEs are obtained
for the $S_{j}$'s.

The time-dependent WKB begins by inserting the
ansatz\cite{pauli,kurt}
\begin{equation}
\label{ansatz2}
 \psi(x,t)=\exp\left[\frac{i}{\hbar}S(x,t)\right],
\end{equation}
into the time-dependent Schr\"odinger equation,
\begin{equation}
i\hbar\partial_{t}\psi=-\frac{\hbar^{2}}{2m}\partial_{xx}\psi+V(x,t)\psi,
\end{equation}
The result is the quantum Hamilton-Jacobi equation\cite{pauli,kurt}
\begin{equation}
\label{st1}
 \partial_{t}S+\frac{1}{2m}(\partial_{x}S)^{2}+V=\frac{i\hbar}{2m}\partial_{xx}S,
\end{equation}
where the LHS of the equation is in the form of the classical
Hamilton-Jacobi equation. Equation (\ref{st1}) is formally exact
since no approximation has been introduced. In TDWKB formulation we
insert into eq.(\ref{st1}) a time-dependent version of
eq.(\ref{expansion1})
\begin{equation}
\label{expansion2}
 S(x,t)=\sum_{j=0}^{\infty}\hbar^{j}S_{j}(x,t).
\end{equation}
The result is
\begin{equation}
\label{power_h}
 \sum_{j=0}^{\infty}\hbar^{j}\partial_{t}S_{j}+\frac{1}{2m}\sum_{j_{1},j_{2}=0}^{\infty}
 \hbar^{j_{1}+j_{2}}\partial_{x}S_{j_{1}}\partial_{x}S_{j_{2}}+V
 =\frac{i}{2m}\sum_{j=0}^{\infty}\hbar^{j+1}\partial_{xx}S_{j}.
\end{equation}
By equating terms having the same powers of $\hbar$ we obtain the
classical Hamilton-Jacobi equation for $S_{0}(x,t)$
\begin{equation}
\label{st2}
\partial_{t}S_{0}+\frac{1}{2m}(\partial_{x}S_{0})^{2}+V=0,
\end{equation}
and equations of motion for $S_{n}(x,t), n\geq1$
\begin{equation}
\label{st3}
\partial_{t}S_{n}+\frac{\partial_{x}S_{0}}{m}\partial_{x}S_{n}=\frac{i}{2m}\partial_{xx}S_{n-1}-
\frac{1}{2m}\sum_{j=1}^{n-1}\partial_{x}S_{j}\cdot\partial_{x}S_{n-j}.
\end{equation}
Conveniently, each equation depends only on lower order terms. The
next step in TDWKB is to convert eqs.(\ref{st2}) and (\ref{st3})
into a set of ODEs by looking at the evolution of $S_0,S_1,\ldots$
along classical trajectories, as described in the next section.

\subsection{Integrating along classical trajectories}
\label{trajectories}

\noindent As we mentioned earlier, the first term in the $\hbar$
power expansion, $S_{0}$, obeys the Hamilton-Jacobi equation
(eq.(\ref{st2})). This equation is an alternative formulation of
Newton's second law of motion in terms of an action field. The
emergence of classical trajectories in the TDWKB equations provides
the incentive to solve eqs.(\ref{st2}) and (\ref{st3}) by
integrating along such trajectories.

The link between the Hamilton-Jacobi equation and classical
trajectories is demonstrated by defining the {\em velocity field}
\begin{equation}
\label{def1}
 v(x,t)\equiv\frac{\partial_{x} S_{0}(x,t)}{m}
\end{equation}
and considering the trajectories defined by
\begin{equation}
\label{dxdt} \frac{dx}{dt}=v(x,t)\  .
\end{equation}
By taking the spatial partial derivative of eq.(\ref{st2}), using
the definition of the Lagrangian time derivative
$\frac{d}{dt}\equiv\partial_{t}+\frac{dx}{dt}\partial_{x}$, and
applying eq.(\ref{def1}) we obtain the equation of motion for the
velocity along a trajectory as Newton's second law
\begin{equation}
\label{dvdt}
 \frac{dv}{dt}=-\frac{\partial_{x}V}{m}.
\end{equation}
Hence, the trajectories defined are simply classical trajectories.

Inserting eq.(\ref{st2}) in the Lagrangian time derivative of
$S_{0}$ yields
\begin{equation}
\label{s0}
 \frac{dS_{0}}{dt}=\partial_{t}S_{0}+\frac{\partial
 S_{0}}{m}\partial_{x}S_{0}=\frac{1}{2}mv^{2}-V,
\end{equation}
where we recognize the equation of motion for the action along a
classical trajectory. Noting that $v$ is a mere dummy variable, we
summarize the equations of motion for the zeroth order term of
TDWKB, $S_{0}$
\begin{eqnarray}
\label{set_s0}
 \frac{dx}{dt}&=&\frac{\partial_{x}S_{0}}{m}, \\ \label{set_s01}
 \frac{d(\partial_{x}S_{0})}{dt}&=&-\partial_{x}V, \\
 \label{set_s02}
 \frac{dS_{0}}{dt}&=&\frac{1}{2m}(\partial_{x}S_{0})^{2}-V.
\end{eqnarray}
We turn to the higher order terms in the series $S_{n}$, $n\geq1$.
Recognizing the LHS of eqs.(\ref{st3}) as the Lagrangian time
derivative of $S_{n}$, we can write
\begin{equation}
\label{st4}
 \frac{dS_{n}}{dt}=\frac{i}{2m}\partial_{xx}S_{n-1}-
\frac{1}{2m}\sum_{j=1}^{n-1}\partial_{x}S_{j}\cdot\partial_{x}S_{n-j}.
\end{equation}
These equations do not constitute a closed set of ODEs since they
depend on partial derivatives such as $\partial_{xx}S_{n}$. We close
the set of equations by deriving equations of motion for the partial
derivatives on the RHS of eq.(\ref{st4}) ($\partial_{xx}S_{n-1}$,
$\partial_{x}S_{j}$ and $\partial_{x}S_{n-j}$). We demonstrate the
process by deriving equations of motion for $S_{1}$ and $S_{2}$.
Inserting $n=1$ in eq.(\ref{st4}) yields
\begin{equation}
\label{dtds1}
 \frac{dS_{1}}{dt}=\frac{i}{2m}\partial_{xx}S_{0}.
\end{equation}
An equation of motion for $\partial_{xx}S_{0}$ is obtained by taking
a second spatial partial derivative of eq.(\ref{st2}),
\begin{equation}
\label{dtdsxx}
 \partial_{xxt}S_{0}+\frac{1}{m}\left[\partial_{x}S_{0}\cdot\partial_{xxx}S_{0}+(\partial_{xx}S_{0})^{2}\right]+\partial_{xx}V=0,
 \end{equation}
and rewriting it as
\begin{equation}
\label{dtds0xx}
 \frac{d(\partial_{xx}S_{0})}{dt}=-\frac{1}{m}(\partial_{xx}S_{0})^{2}-\partial_{xx}V.
 \end{equation}
This equation is derived in reference \cite{boiron} by a cumbersome
finite difference scheme. It is equivalent to eq.(2.9d) of reference
\cite{huber2} where the equation appears in the context of GGWPD.
Note that an equation of motion for any order of spatial derivatives
of $S_{0}$ can be derived in a similar fashion by taking consecutive
spatial derivatives of eq.(\ref{dtdsxx}) and then grouping the
Lagrangian time derivative terms. Equations (\ref{dtds1}) and
(\ref{dtds0xx}) provide a closed set of equations of motion for
$S_{1}$.

Inserting $n=2$ into eq.(\ref{st4}) yields
\begin{equation}
\label{dtds2}
 \frac{dS_{2}}{dt}=\frac{i}{2m}\partial_{xx}S_{1}-\frac{1}{2m}(\partial_{x}S_{1})^2.
\end{equation}
The equations of motion for $\partial_{x}S_{1}$ and
$\partial_{xx}S_{1}$ are obtained by first inserting $n=1$ in
eq.(\ref{st3}). We then derive two equations by taking a first and a
second spatial partial derivative of the result. By grouping the
Lagrangian time derivatives of $\partial_{x}S_{1}$ and
$\partial_{xx}S_{1}$ in each of the two equations separately we
obtain
\begin{eqnarray}
\label{st51}
\frac{d(\partial_{x}S_{1})}{dt}&=&\frac{i}{2m}\partial_{xxx}S_{0}-\frac{1}{m}\partial_{x}S_{1}\cdot\partial_{xx}S_{0},
\\ \nonumber
\frac{d(\partial_{xx}S_{1})}{dt}&=&\frac{i}{2m}\partial_{xxxx}S_{0}-\frac{1}{m}\partial_{x}S_{1}\cdot\partial_{xxx}S_{0}
-\frac{2}{m}\partial_{xx}S_{1}\cdot\partial_{xx}S_{0}.
\end{eqnarray}
The last equations depend in turn on $\partial_{xxx}S_{0}$ and
$\partial_{xxxx}S_{0}$. As mentioned earlier, the equation of motion
for these terms can be obtained by additional spatial derivatives of
eq.(\ref{dtdsxx}), a process that yields
\begin{eqnarray}
\label{st52}
 \frac{d(\partial_{xxx}S_{0})}{dt}&=&-\frac{3}{m}\partial_{xx}S_{0}\cdot\partial_{xxx}S_{0}-\partial_{xxx}V,\\
 \nonumber
 \frac{d(\partial_{xxxx}S_{0})}{dt}&=&-\frac{1}{m}\left[4\partial_{xx}S_{0}\cdot\partial_{xxxx}S_{0}
 -3(\partial_{xxx}S_{0})^{2}\right]-\partial_{xxxx}V.
\end{eqnarray}
Equations (\ref{dtds0xx}) and (\ref{dtds2})-(\ref{st52}) provide a
closed set of equations of motion for $S_{2}$. The scheme we
described for $S_{1}$ and $S_{2}$ can be extended to any of the
higher order terms in the expansion. Note that incorporating higher
order terms $S_{n}$ in the TDWKB approximation does not affect the
classical trajectories associated with $S_{0}$, defined by
eqs.(\ref{set_s0}) and (\ref{set_s01}). We now turn to the source of
the distinction between conventional TDWKB and CTM.


\subsection{Initial conditions and complex classical trajectories}

\label{initial} \noindent In conventional TDWKB the initial
wavefunction is ``divided" between $S_{0}(x,0)$ and $S_{1}(x,0)$
\begin{equation}
\psi(x,0)=A(x)\exp[i\phi(x)]=\exp\left[\frac{i}{\hbar}S_{0}(x,0)+S_{1}(x,0)\right],
\end{equation}
where $A(x)$ and $\phi(x)$ are the initial amplitude and phase
respectively, both taken to be real. The phase is related to the
zero-order term $S_{0}$ and the amplitude to the first-order
correction term $S_{1}$ according to
\begin{equation}
\label{last}
 S_{0}(x,0)=\hbar\phi(x), \ \ \ S_{1}(x,0)=-i\ln[A(x)],
\end{equation}
and $S_{n}(x,0)=0$ for $n\geq2$. Note that the initial conditions
specified by eqs.(\ref{last}) yield classical trajectories that
propagate on the \textit{real} axis since $S_{0}$ and its spatial
derivatives are real quantities (see eqs.(\ref{set_s0}) and
(\ref{set_s01})). In contrast, in CTM the amplitude and phase are
treated on an equal footing with far-reaching consequences. The
initial wavefunction is specified by $S_{0}(x,0)$
\begin{equation}
\label{last2}
S_{0}(x,0)=-i\hbar\ln[\psi(x,0)], \ S_{n}(x,0)=0, \
n\geq1.
\end{equation}
Since $S_{0}$ is generally complex and since the initial velocity
$v(x,0)\equiv\partial_{x}S_{0}(x,0)/m$, the trajectories propagate
in the \textit{complex plane} even if the initial positions are on
the real axis ($\Im[x(0)]=0$). This observation requires us to look
at the analytic continuation of the wavefunction in the complex
plane and find ways to extract the wavefunction on the real axis.

\subsection{Complex root search and superposition}
\label{super}

\noindent One of the benefits of conventional TDWKB and CTM compared
with BOMCA, is that the trajectories obey the classical equations of
motion and are independent of the order of the phase expansion we
incorporate in the final wavefunction. But the fact still remains
that for an arbitrary initial position $x(0)\in \mathbb{C}$ and an
arbitrary final propagation time $t_{f}$ the final position
$x(t_{f})$ is complex and yields an ``analytically continued"
wavefunction at $x(t_{f})$
\begin{equation}
\psi[x(t_{f}),t_{f}]\approx\exp\left\{\frac{i}{\hbar}\sum_{j=0}^{N}\hbar^{j}S_{j}[x(t_{f}),t_{f}]\right\},
\end{equation}
where the non-negative integer $N$ is the order of the
approximation. References \cite{boiron,goldfarb,huber2} include
discussions of root search algorithms for the derivation of initial
positions that reach the \textit{real} axis at a given time. We will
not describe all the details here but will just state the central
idea. The complex root search exploits the assumption that the
mapping $x(0)\mapsto x(t_{f})$ is analytic. This property allows for
an iterative process that detects the initial positions that
correspond to real final positions. As demonstrated in references
\cite{huber1,huber2} and in section \ref{frt_ord}, for an arbitrary
potential and final time, the mapping is only \textit{locally}
analytic. Generally, more than one initial position ends at a final
position (whether real or complex). This makes the search for
trajectories that end on the real axis more complicated but it has
an important advantage in terms of interference effects.

Our main observation is that the contribution of multiple
trajectories in CTM can accumulate to an interference pattern. For
simplicity we make the following assumption. Suppose that $L$
trajectories end at final time $t_{f}$ on real position $x(t_{f})$
Then the final wavefunction is approximated by a superposition of
contributions
\begin{equation}
\label{sup}
\psi[x(t_{f}),t_{f}]\approx\sum_{l=1}^{L}\exp\left\{\frac{i}{\hbar}S^{l}\left[x(t_{f}),t_{f}\right]\right\},
\end{equation}
where each trajectory (denoted by the index $l$) is associated with
a phase $S^{l}[x(t_{f}),t_{f}]$
\begin{equation}
S^{l}[x(t_{f}),t_{f}]=\sum_{j=0}^{N}\hbar^{j}S_{j}^{l}[x(t_{f}),t_{f}],
\end{equation}
that is calculated by the CTM equations of motion. In section
\ref{num} we show that this assumption is too simplified and does
not hold at all times and all positions. For example, for positions
associated with a tunneling part of the wavefunction, only one of
the multiple trajectories should be taken into account. A partial
discussion on the superposition of contributions from complex
trajectories appears in reference \cite{huber2} in the GGWPD
context. In a forthcoming paper \cite{newgst} we will explain an
alternative derivation of the CTM in which the need to include
multiple trajectories for certain times and positions becomes
apparent

%
%

\section{Numerical Results}
\label{num}

\noindent In this section we examine numerically the CTM formulation
allowing for the superposition of complex trajectories. For ready
comparison the physical system we choose is identical to the one
studied by Boiron and Lombardi (reference \cite{boiron} section
IVB). The potential considered is a quartic double-well
\begin{equation}
V(x)=1.25\times10^{-4}(x^{4}-400x^{2}).
\end{equation}
The initial wavefunction is a Gaussian wavepacket
\begin{equation}
 \psi(x,0)=\exp\left[-\alpha_{0}
 (x-x_{c})^{2}+\frac{i}{\hbar}p_{c}(x-x_{c})+\frac{i}{\hbar}\gamma_{0}\right],
\end{equation}
where $\alpha_{0}=1$, $x_{c}=0$, $p_{c}=5$, $\gamma_{0}=-\frac{i
\hbar}{4}\ln(\frac{2\alpha_{0}}{\pi})$ and we take $m=\hbar=1$ (all
quantities are given in atomic units). The initial conditions for
the terms in the $\hbar$ power-expansion of the phase are
\begin{eqnarray}
 S_{0}(x,0)&=&i\alpha_{0}\hbar(x-x_{c})^{2}+p_{c}(x-x_{c})+\gamma_{0}=ix^{2}+5x+\gamma_{0},
 \\ \label{moment0}
 \partial_{x} S_{0}(x,0)&=&2i\alpha_{0}\hbar(x-x_{c})+p_{c}=2ix+5, \\
 \partial_{xx} S_{0}(x,0)&=&2i\alpha_{0}\hbar=2i, \\
 \partial_{x}^{j} S_{0}(x,0)&=&0, \ j\geq 3, \\
 \partial_{x}^{j} S_{k}(x,0)&=&0, \ j\geq 0, \ k\geq 1,
\end{eqnarray}
where $\partial_{x}^{j} S_{k}\equiv\frac{\partial^{j}S_{k}}{\partial
x^{j}}$.

In section \ref{frt_ord} we analyze the first order approximation of
CTM ($N=1,\ S= S_{0}+\hbar S_{1}$) and the properties of the
trajectories. Section \ref{sec_ord} is dedicated to the next order
of the approximation ($N=2,\ S= S_{0}+\hbar S_{1}+\hbar^{2}S_{2}$).
We omit an analysis of $N=0$ since it is well presented in reference
\cite{boiron} and only yields poor results.

\subsection{First Order approximation, $N=1$}
\label{frt_ord}

\begin{figure}
\begin{center}
\epsfxsize=9 cm \epsfbox{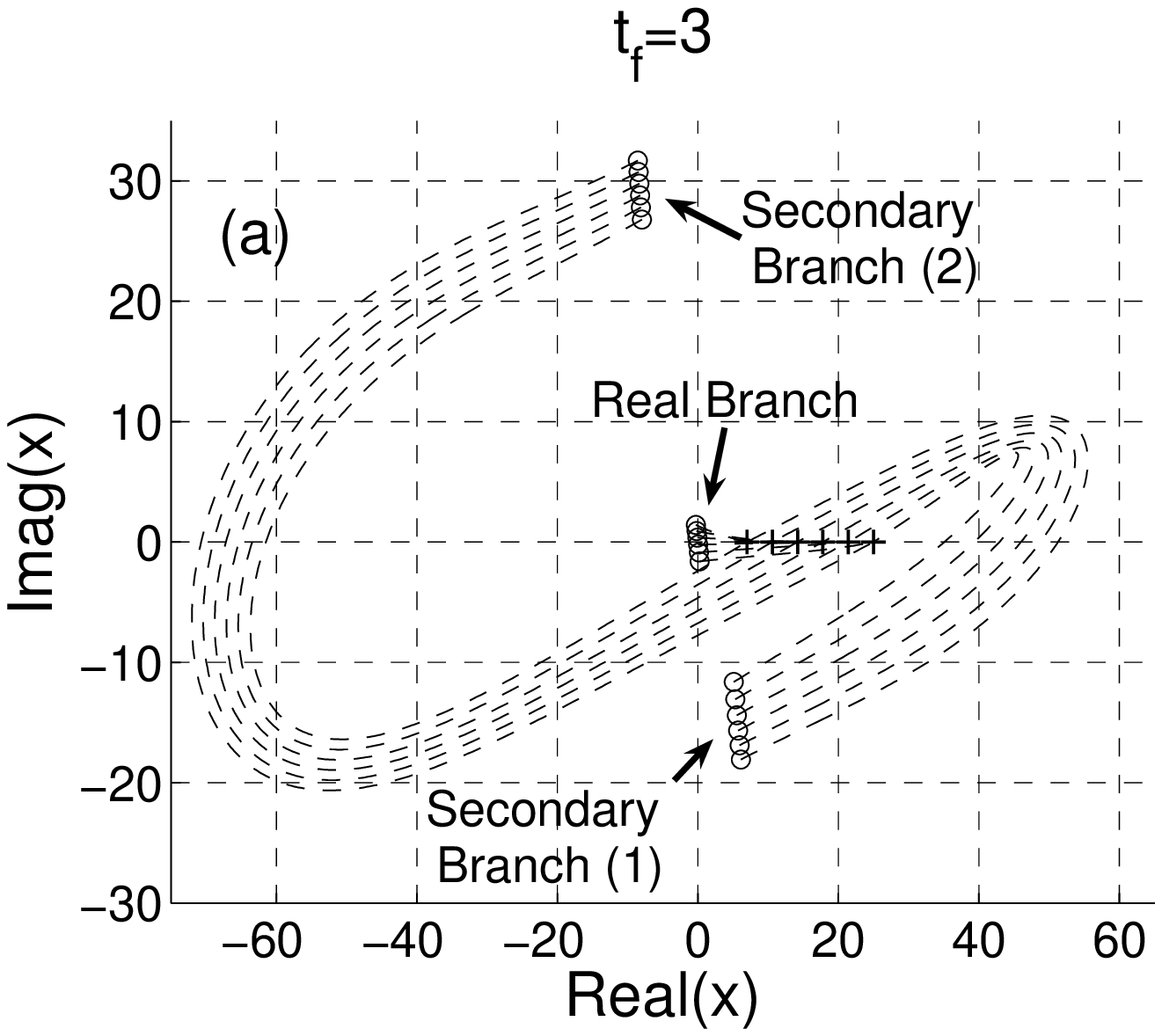} 
\epsfxsize=9 cm \epsfbox{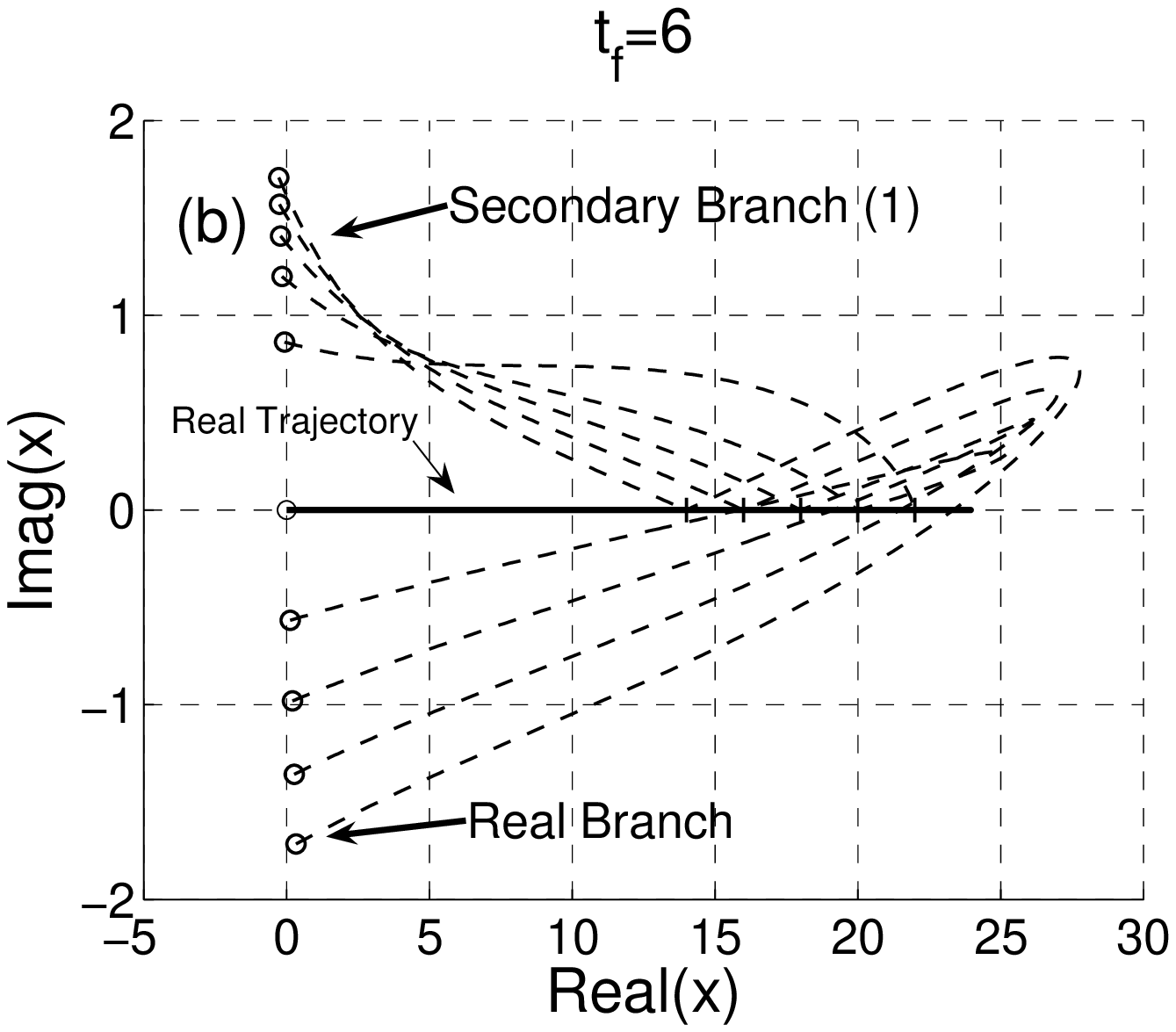} 
\end{center}
\caption{\label{tra} Complex classical trajectories with initial
positions marked as circles and \textit{real} final positions
(marked as pluses) at (a) $t_{f}=3$  and (b) $t_{f}=6$. The
trajectories arise from an initial Gaussian wavepacket propagating
in a quartic double-well potential. The Gaussian is centered at
$x=0$ and has positive initial momentum (the physical parameters are
given in the text). In plot (a) we demonstrate that each final
position arises from three initial positions. The initial positions
are divided into a real branch and two secondary branches. The real
branch is defined as incorporating a trajectory that remains on the
real axis at all times. The real trajectory is specifically
indicated in plot (b).}
\end{figure}

\noindent The first order approximation of CTM requires the solution
of eqs.(\ref{set_s0}), (\ref{set_s01}), (\ref{set_s02}),
(\ref{dtds1}) and (\ref{dtds0xx}). The first two equations define
the complex classical trajectories and the next three equations
yield $S_{0}$ and $S_{1}$. We start by analyzing the complex
classical trajectories. As mentioned above, the mapping $x(0)\mapsto
x(t_{f})$ is not one-to-one. For the quartic potential, we found
that three initial positions are mapped to every real final position
at $t_{f}>0$. For short time scales this observation can be
supported analytically. For general potentials or for longer time
scales than we present here, more than three initial positions might
lead to the same final position\cite{huber2,goldfarb2}. In figures
\ref{tra}(a) and \ref{tra}(b) we plot complex classical trajectories
for $t_{f}=3$ and $t_{f}=6$ respectively. The \textit{initial
positions} of the trajectories can be divided into three groups
referred to as \textit{branches}\cite{huber2}. One group of the
initial positions is called the real branch and the other groups are
called the secondary branches. The real branch is characterized by
the property that it includes the initial position of a trajectory
that propagates \textit{solely} on the real axis. We refer to this
trajectory as the real trajectory. It can be readily verified that
for a Gaussian initial wavefunction there is only a single real
trajectory that initiates at $x(0)=x_{c}$ (see eqs.(\ref{set_s01}),
eq.(\ref{set_s0}) and (\ref{moment0})). In fig.\ref{tra}(b) we
depict the real trajectory explicitly. The secondary branches are
defined simply as the groups of initial positions that do not belong
to the real branch. Generally, the branches might be infinitely long
curves in the complex plane. We will use the term branches to refer
to the locus of \textit{initial} positions that leads to final
positions where the wavefunction is significantly different from
zero. Hence, the branches are curves of \textit{finite} length in
the complex plane, although clearly there is some arbitrariness to
their length.

In fig.\ref{tra}(a) we see that at short time scales the secondary
branches are centered far from neighborhood of the real axis. We can
show analytically that for small times $t_f$ the initial positions
that comprise the real branch obey $|x(0)|= O({t_{f}})$ whereas the
secondary branches obey $|x(0)|=O(\frac{1}{t_{f}})$. Note that the
linear dependence of the initial momentum on position
(eq.(\ref{moment0})) allows trajectories with initial positions far
from the real axis to reach a real final position in a short time.
Unlike the secondary branches, the real branch is centered in the
vicinity of the real axis at all times. The initial position
$x(0)=x_{c}$ is a fixed point of the real branch and prevents the
real branch (recall that this is the locus of \textit{initial}
positions) from ``straying" from the neighborhood of the real axis
as the final time $t_f$ is increased. At intermediate times (time
scales comparable to the time of the collision of the wavefunction
with the barrier, $4\lesssim t_{f}\lesssim7$) secondary branch (1)
reaches the vicinity of the real axis (fig.\ref{tra}(b)) and at
longer time scales it continues in the direction of the positive
imaginary axis. As we demonstrate below, the proximity of secondary
branch (1) to the real axis is closely related to the size of its
contribution to the final form of the wavefunction and its role in
interference effects. Secondary branch (2) does not reach the
vicinity of the real axis for any of the time scales specified
below. The contribution of this branch to the absolute value of the
final wavefunction (eq.(\ref{sup})) is negligible (in the order of
$10^{-35}$). Hence, from here on we ignore secondary branch (2) and
refer to secondary branch (1) as \textit{the} secondary branch.

\begin{figure}
\begin{center}
\epsfxsize=8.0 cm \epsfbox{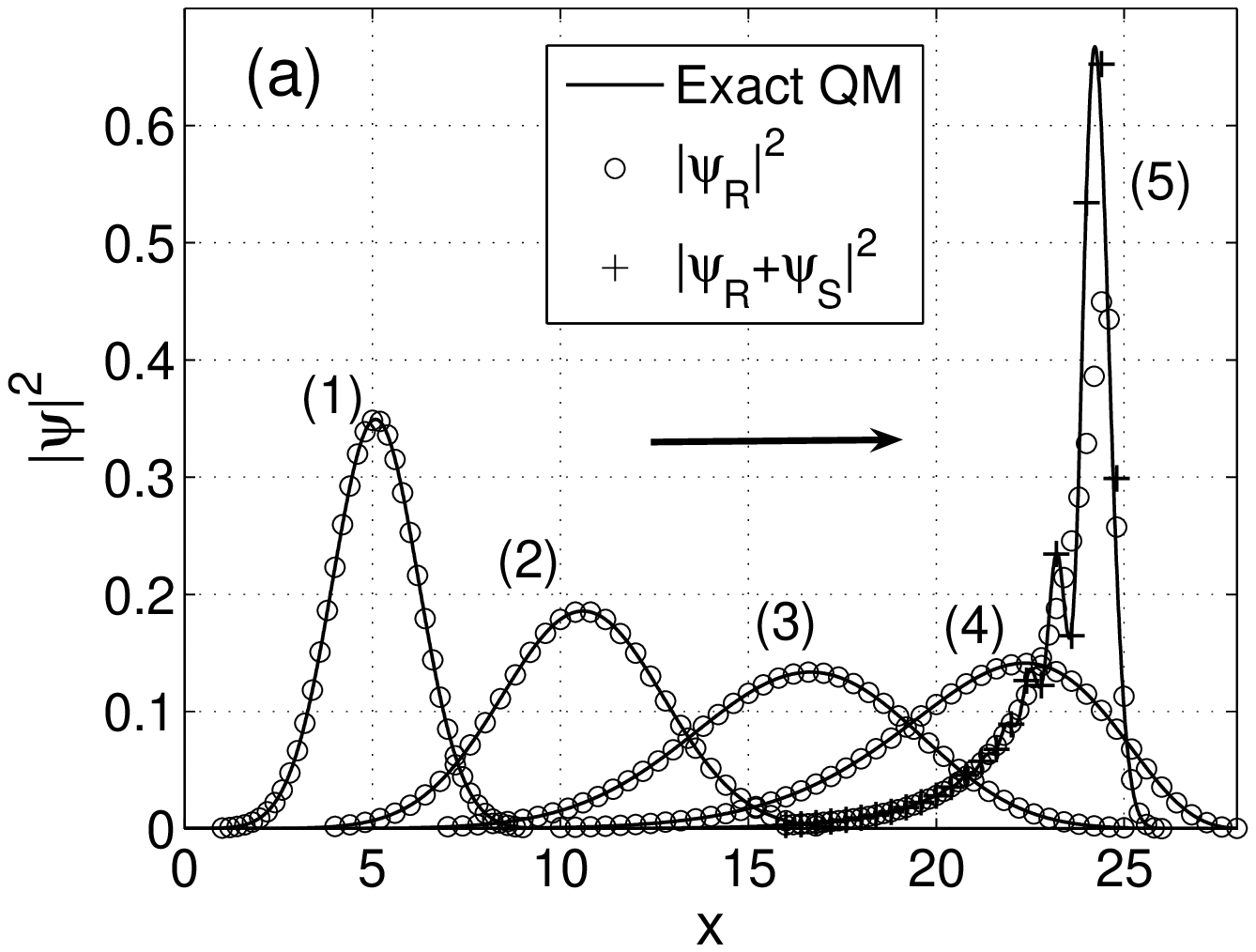} 
\epsfxsize=8.0 cm \epsfbox{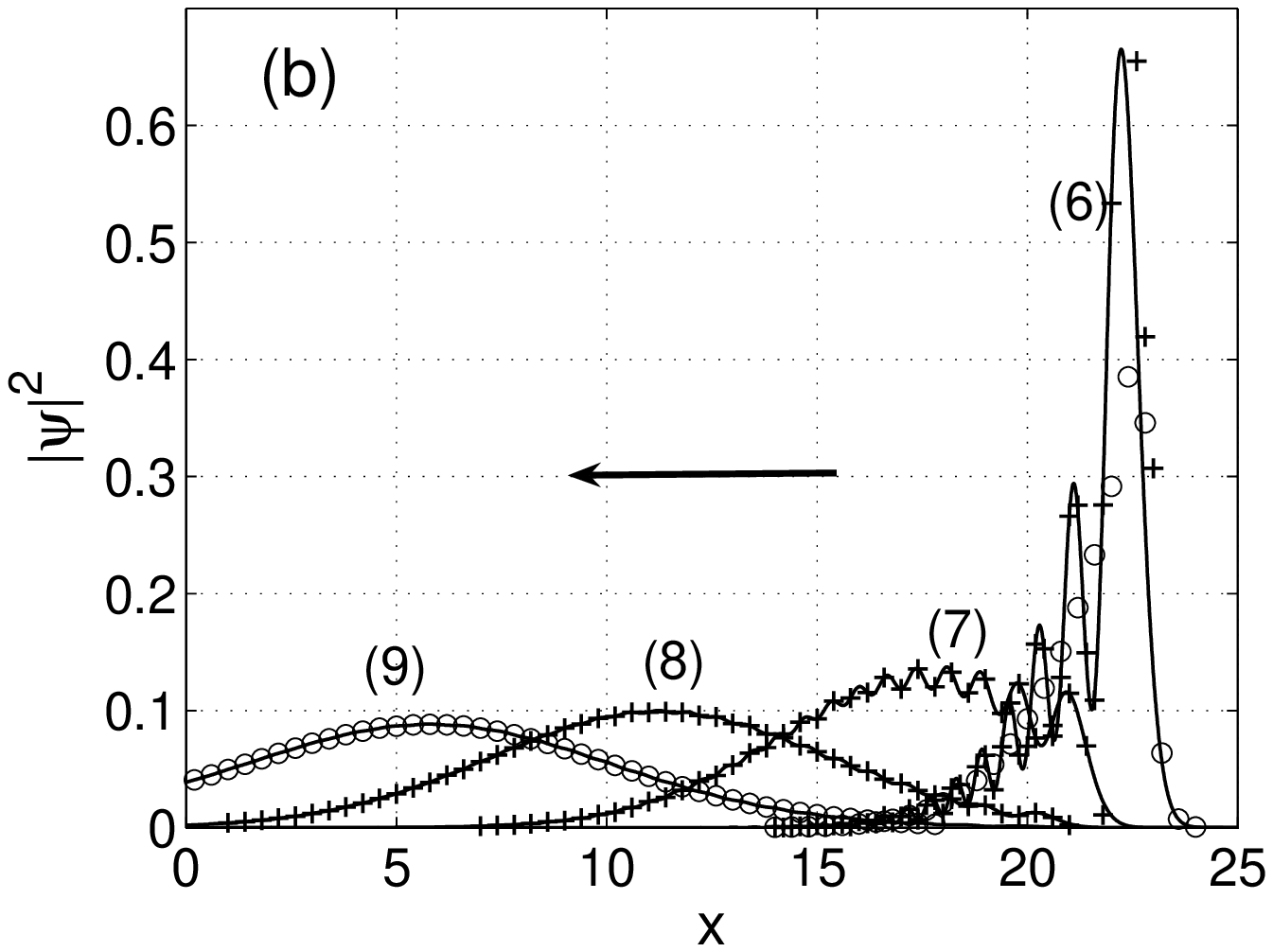} 
\epsfxsize=8.5 cm \epsfbox{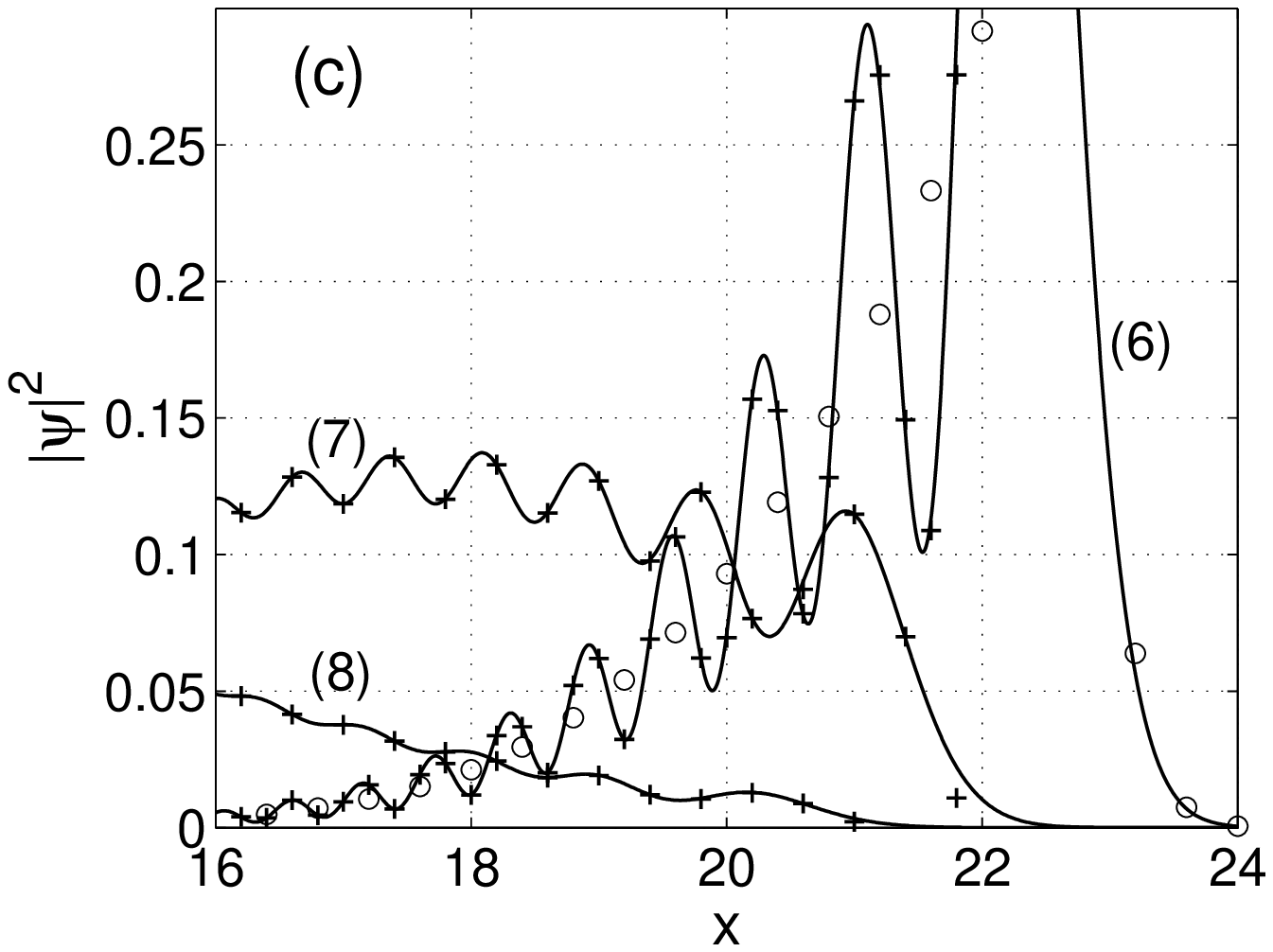} 
\end{center}
\caption{\label{psi1} A comparison between the exact quantum
wavefunction and CTM ($N=1$) with a two-branch superposition. The
comparison is at a series of final propagation times specified by
the numbers in the parentheses. The plots arise from an initial
Gaussian wavepacket centered at $x=0$ with a positive average
momentum, propagating in a quartic double-well potential (the
parameters are given in the text). (a) Initially right-propagating
wavefunction; (b) the reflected wavefunction; (c) a zoom on a
section of (b). For $t_{f}=5$ in (a) and $t_{f}=6$ in (b) and (c) we
plot the results both for just the real branch
$|\psi_{\textrm{R}}|^{2}$ and for the combination of branches
$|\psi_{\textrm{R}}+\psi_{\textrm{S}}|^{2}$. The interference
pattern obtained by superposing the contributions is clearly
observed.}
\end{figure}

As we mentioned in section \ref{super}, the existence of more than
one branch motivates the attempt to superpose the contributions of
the real branch and secondary branch in the final wavefunction
\begin{eqnarray}
\psi[x(t_{f}),t_{f}]=\psi_{\textrm{R}}+\psi_{\textrm{S}}; \ \ \
\psi_{\textrm{R}}=\exp\left[\frac{i}{\hbar}S_{\textrm{Real}}\right],
\
\psi_{\textrm{S}}=\exp\left[\frac{i}{\hbar}S_{\textrm{Sec}}\right],
\end{eqnarray}
where $S_{\textrm{Real}}$ and $\psi_{\textrm{R}}$ are the phase and
wavefunction associated with the real branch, and $S_{\textrm{Sec}}$
and $\psi_{\textrm{S}}$ correspond to the secondary branch. In
figures \ref{psi1}(a), \ref{psi1}(b) and \ref{psi1}(c) we compare
the exact wavefunction with the numerical results obtained by
applying CTM using a two-branch superposition. The figures indicate
that when the wavefunction does not exhibit oscillations, the
contribution of the real branch is sufficient to obtain a good
approximation to the wavefunction. But at intermediate times, when
the wavefunction exhibits interference effects, the contribution of
both branches must be included. This last last observation applies
in the spatial range up to the classical turning point
($x\simeq24$), beyond which the combined contribution diverges from
the exact result.

\begin{figure}[t]
\begin{center}
\epsfxsize=9 cm \epsfbox{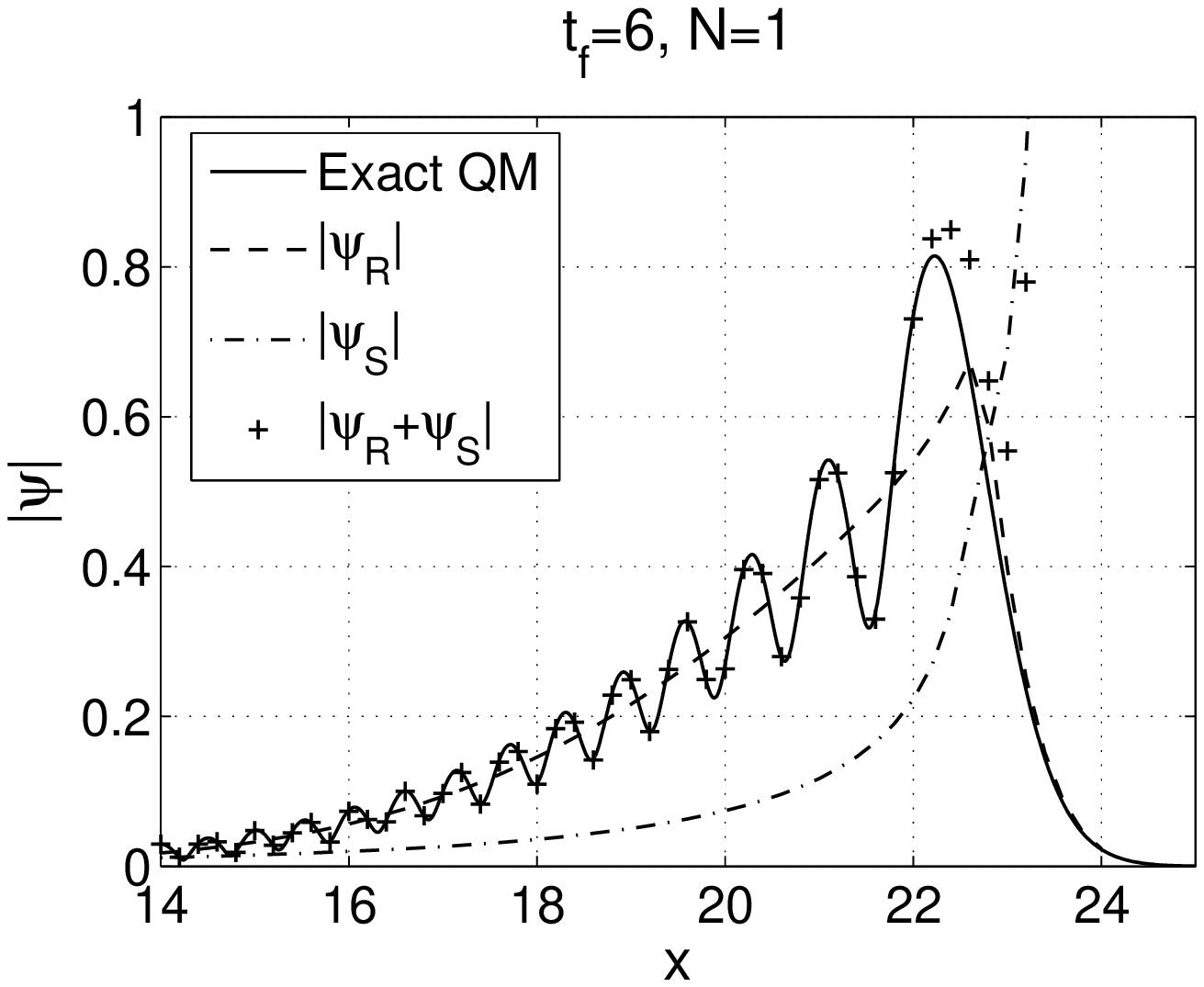} 
\end{center}
\caption{\label{prop6_1} CTM approximation at $t_{f}=6$ for $N=1$.
The contributions of each branch to the wavefunction are depicted by
plotting $|\psi_{\textrm{R}}|$, $|\psi_{\textrm{S}}|$ and
$|\psi_{\textrm{R}}+\psi_{\textrm{S}}|$. Note the exponential
increase of $\psi_{\textrm{S}}$ begins around $x\simeq22$. For
$x\simeq23$; we discard the contribution of the secondary branch and
include just the real branch, leading to a discontinuity of the CTM
approximation.}
\end{figure}

We turn to a closer inspection of this divergence. In
fig.\ref{prop6_1} we plot the contribution at $t_{f}=6$ of each
individual branch and their superposition. Starting from the
vicinity of $x\simeq22$, we observe an exponential increase of
$\psi_{\textrm{S}}$. For $x\gtrsim23$ we have a discontinuity of the
approximation, as we discard the contribution of the secondary
branch and include just the real branch. A description of this
divergence appears in reference \cite{huber2} in the context of the
GGWPD formulation.

It is interesting to compare the time-dependence of the real and
secondary branch contributions to the final approximation. A
qualitative measure of the contribution of each branch is given by
the imaginary part of the phase since
\begin{equation}
\label{weight}
|\psi_{\textrm{R}}|=\left|\exp\left(\frac{i}{\hbar}S_{\textrm{Real}}\right)\right|=\exp\left[-\frac{\Im(S_{\textrm{Real}})}{\hbar}\right],
\end{equation}
and a similar relation applies for $\psi_{\textrm{S}}$ and
$\Im(S_{\textrm{Sec}})$. In figures \ref{imag_s}(a) and
\ref{imag_s}(b) we plot $\Im(S_{\textrm{Real}})$ and
$\Im(S_{\textrm{Sec}})$ respectively for a series of final
propagation times. We see that the secondary branch has a
significant magnitude only at intermediate times. This observation
coincides well with the need to include the contribution of the
secondary branch to the final wavefunction only at these times. The
exponential growth of $\psi_{\textrm{S}}$ that is observed in
fig.\ref{prop6_1} is also apparent in fig.\ref{imag_s}(b), in the
negative parts of the graphs for $t_{f}=5$ and $t_{f}=6$. The
divergent magnitude of $\psi_{\textrm{S}}$ is in contrast to the
finite magnitude of $\psi_{\textrm{R}}$ that is observed in
fig.\ref{imag_s}(a). A discontinuity in the derivative of
$\Im(S_{\textrm{R}})$ at $t_{f}=5$ and $t_{f}=6$ is also observed.
This discontinuity appears slightly prior to the points where the
contribution of the secondary branch begins to diverge.
\begin{figure}
\begin{center}
\epsfxsize=9.5 cm \epsfbox{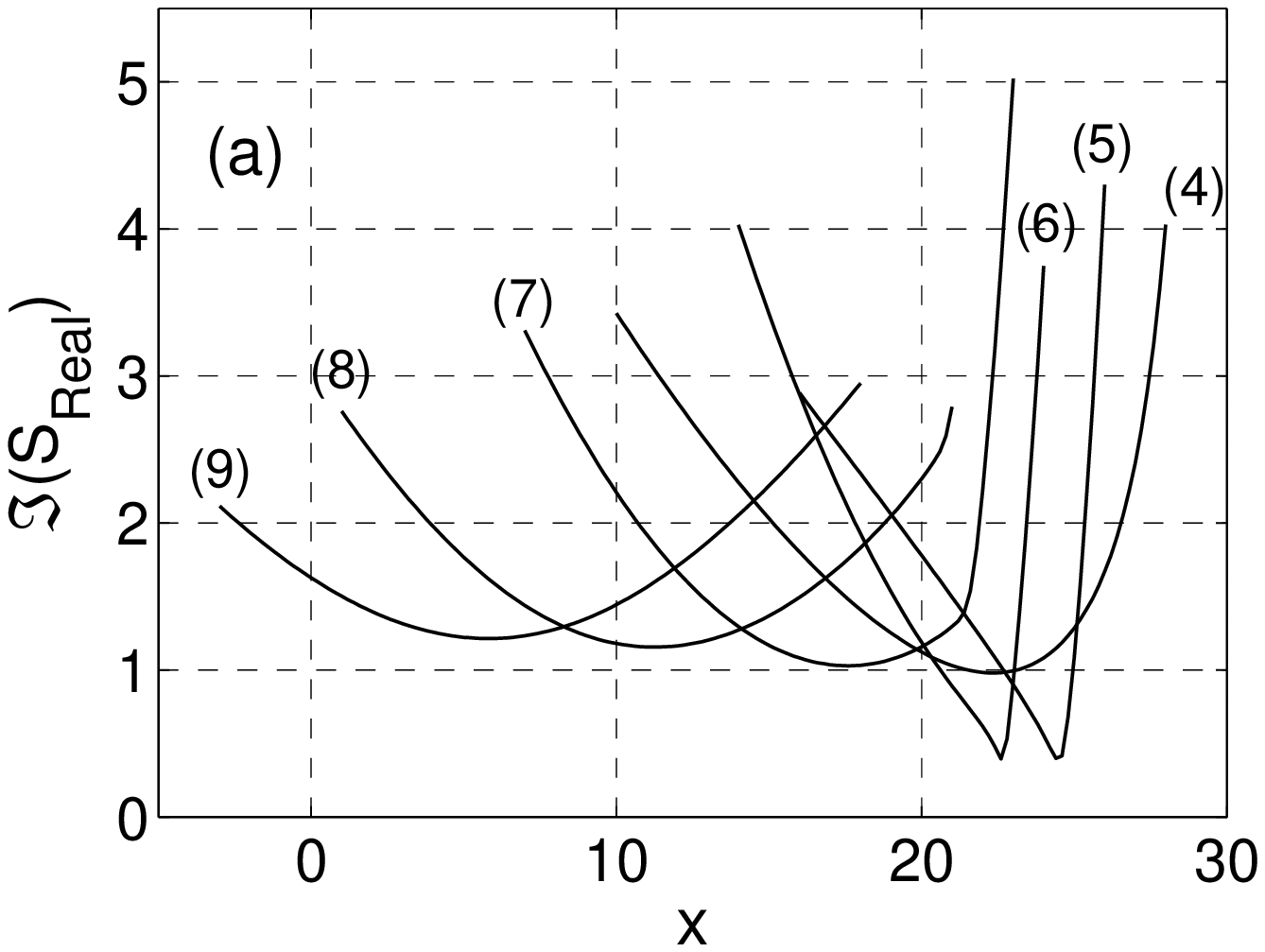} 
\epsfxsize=9.5 cm \epsfbox{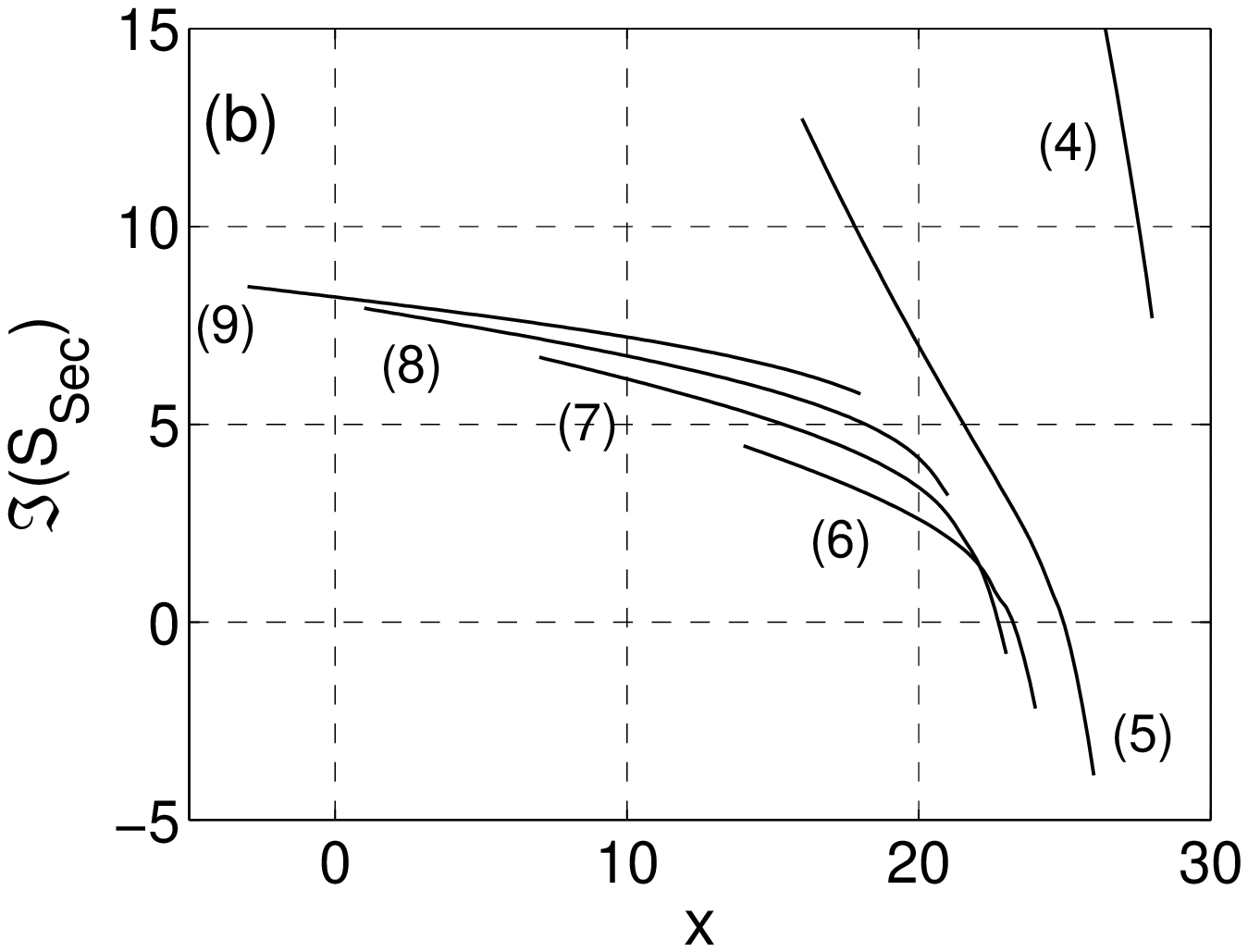} 
\end{center}
\caption{\label{imag_s} (a)  $\Im(S_{\textrm{Real}})$ and (b)
$\Im(S_{\textrm{Sec}})$  are depicted at a series of final
propagation times (given in the parentheses). The results are
limited to the spatial interval for which the absolute value of the
exact wavefunction is significantly larger than zero. The imaginary
part of the phase allows for a qualitative estimate of the
contribution of each branch to the probability
$|\psi_{\textrm{R}}+\psi_{\textrm{S}}|^{2}$, see eq.(\ref{weight}).
Figure \label{imag_s}(b) shows that $\Im(S_{\textrm{Sec}})$ drops
below $\sim2$ only for a finite interval of intermediate times.
Therefore only for this range of times does the secondary branch
makes a significant contribution to the wavefunction.}
\end{figure}

\begin{figure}
\begin{center}
\epsfxsize=9.5 cm \epsfbox{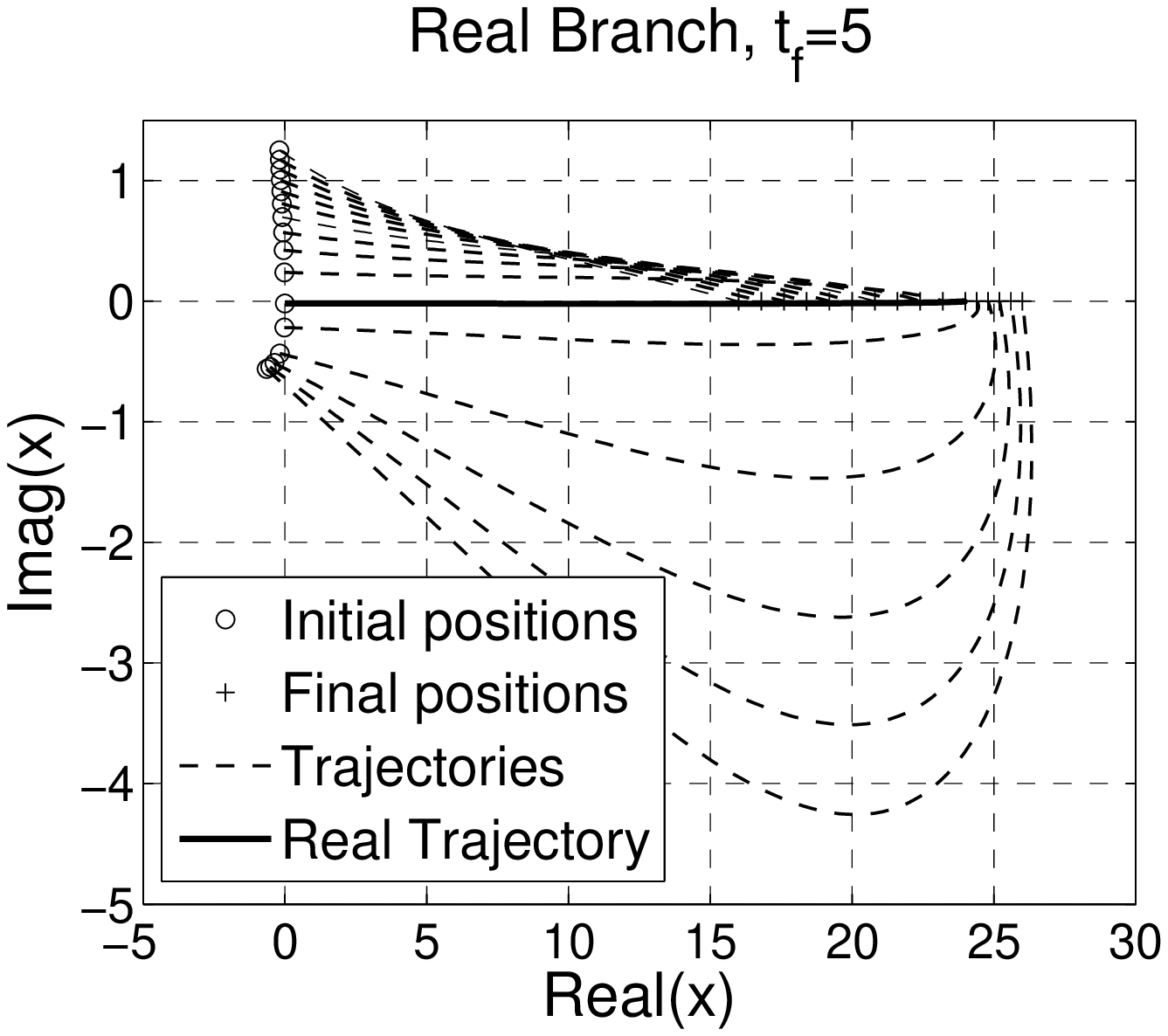} 
\end{center}
\caption{\label{real_tra_t5} Complex classical trajectories that
correspond to the \textit{real branch} at $t_{f}=5$. The real
trajectory acts as a boundary between two ``regimes" of the complex
trajectories. Initial positions with $\Im[x(0)]<0$ seem to be go
past the potential wall. The two ``regimes" can be related to the
singular behavior of the derivative of $\Im(S_{Real })$ at
intermediate times (fig.\ref{imag_s}(a)).}
\end{figure}

A close inspection of the complex trajectories at $t_{f}=5$ and
$t_{f}=6$, reveals an interesting property of the real trajectory:
the real trajectory acts as a boundary between two ``regimes" of
complex trajectories comprising from the real branch. This can be
seen in fig.\ref{real_tra_t5}, where the trajectories that initiate
from $\Im[x(0)]>0$ are seen to reach the real $x$-axis at values
lower than the real trajectory while trajectories with $\Im[x(0)]<0$
seem to go past the barrier and reach the real $x$-axis at values
higher than the real trajectory. These two regimes correspond to the
two legs of the ``v"-shaped graph of $\Im(S_{\textrm{Real}})$ in
fig.\ref{imag_s}(a): the trajectories arising from initial positions
with $\Im[x(0)]>0$ correspond to the left leg of the ``v" while
trajectories with $\Im[x(0)]<0$ correspond to the right leg of the
``v".

\subsection{Second Order approximation, $N=2$}
\label{sec_ord}

\begin{figure}
\begin{center}
\epsfxsize=9 cm \epsfbox{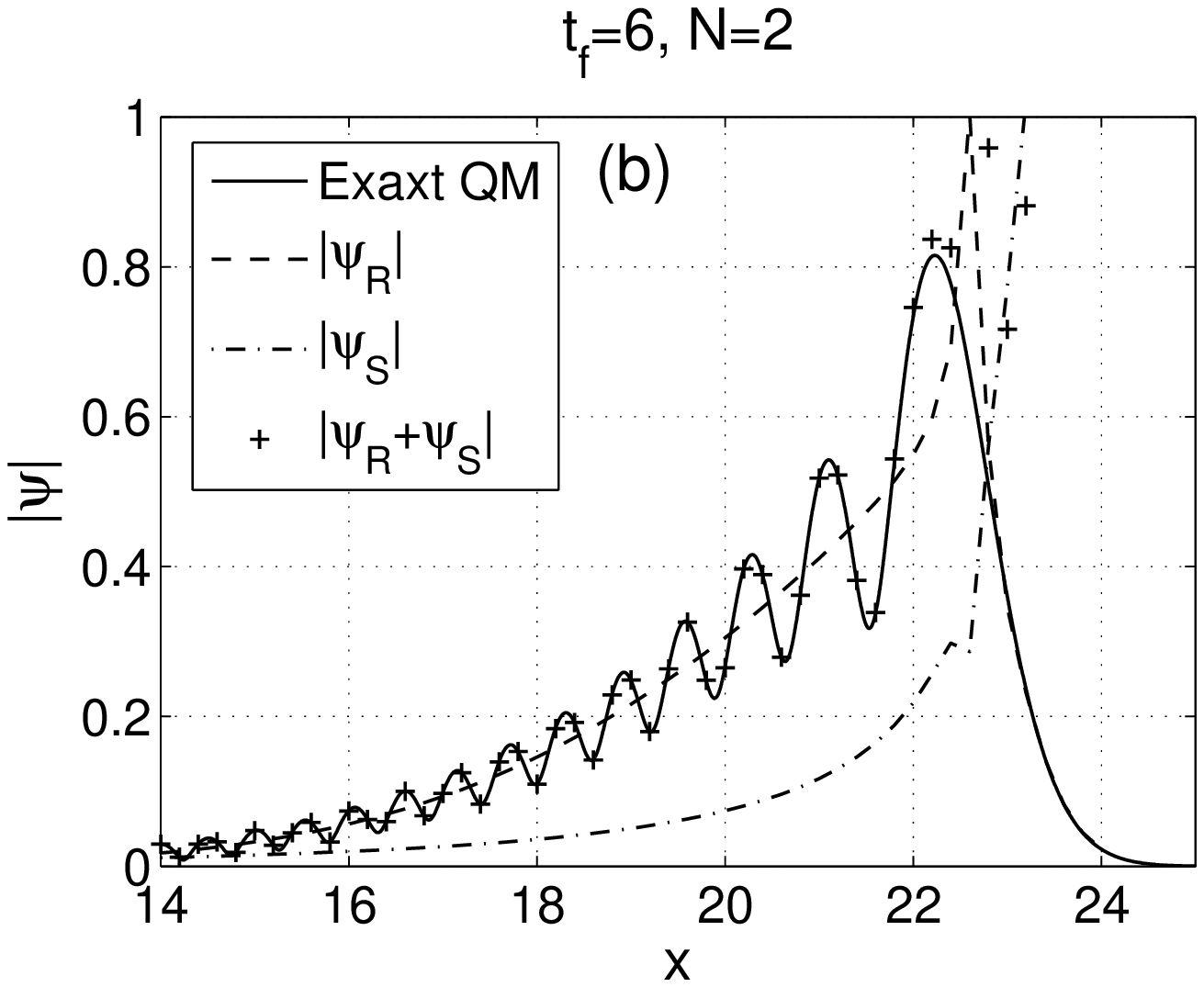} 
\epsfxsize=9 cm \epsfbox{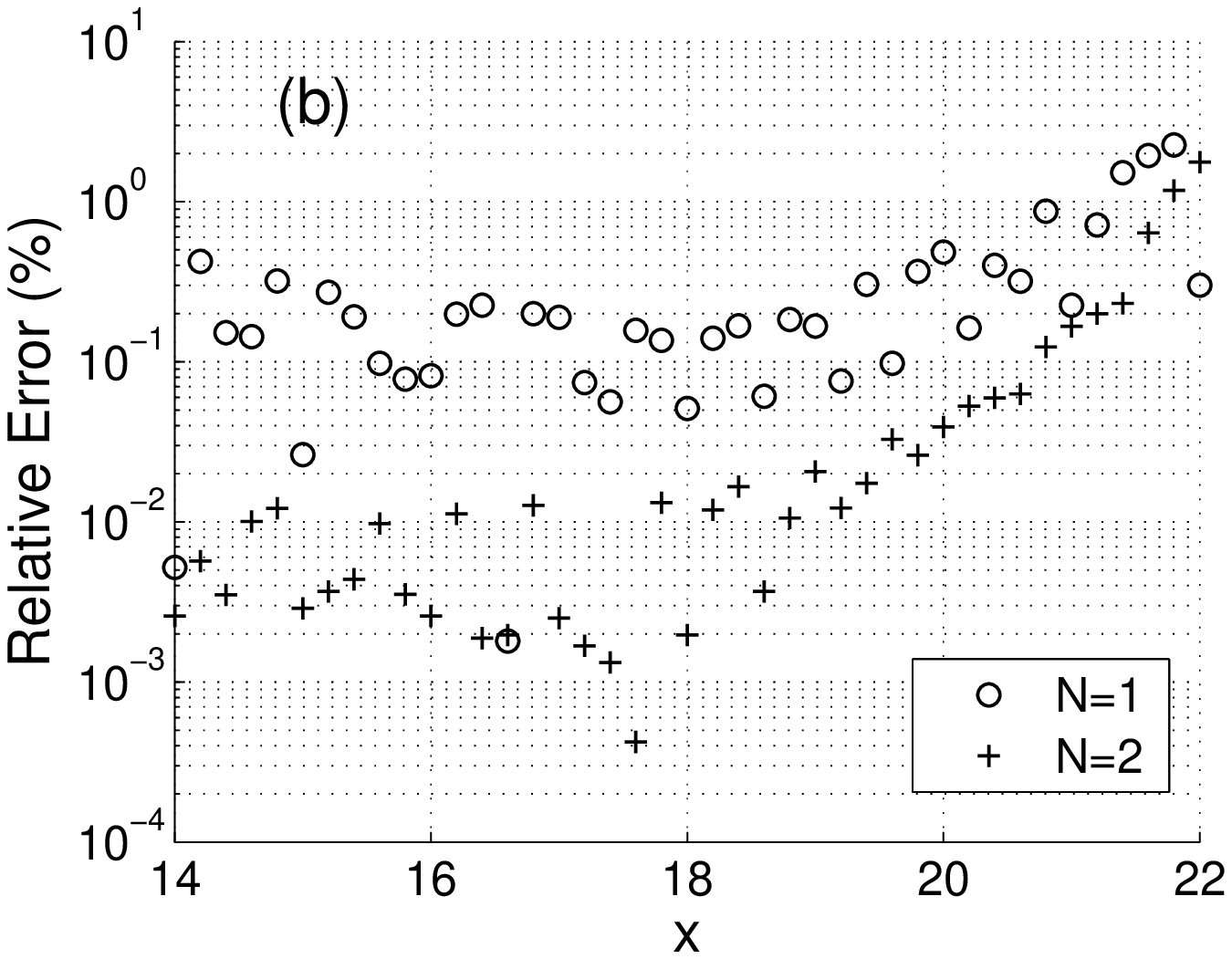} 
\end{center}
\caption{\label{prop6} (a) The second order ($N=2$) CTM
approximation is depicted for $t_{f}=6$. A discontinuity appears at
$x\simeq22.5$ for both $\psi_{\textrm{R}}$ and $\psi_{\textrm{S}}$.
(b) The relative error between the absolute value of the exact
quantum wavefunction and the CTM approximation for $N=1$ and $N=2$,
based on the data in fig.\ref{prop6_1} and fig.\ref{prop6}(a). A
comparison of the relative errors indicates a clear improvement when
taking an additional order in the CTM approximation.}
\end{figure}

\noindent In this section we analyze the effect of incorporating
$S_{2}$ in the CTM approximation. In addition to the five equations
that are needed for obtaining the complex trajectories, $S_{0}$ and
$S_{1}$, we need to solve eqs.(\ref{dtds2}), (\ref{st51}) and
(\ref{st52}). In fig.\ref{prop6}(a) we depict the approximate
wavefunction for $N=2$ at $t_{f}=6$. Comparing the $N=2$ result with
the $N=1$ result plotted in fig.\ref{prop6_1}, we conclude that
other than an interval in the neighborhood of $x\simeq22.5$, the
$N=2$ result (dashed line) lies on top of the exact result (solid
line) and is significantly better then the $N=1$ result. For
$x\gtrsim23$, where we incorporate solely the real branch
contribution, the improvement in the approximation is graphically
evident from the plots. For $x\leq22$ we calculated the relative
error between the absolute value of the approximations and the exact
wavefunction using all the data points depicted in
figs.\ref{prop6_1} and \ref{prop6}(a). The results are presented in
fig.\ref{prop6}(b). For $N=1$  the mean relative error is $ 0.34$\%
while for $N=2$ the mean relative error is $0.11$\%. We see that the
approximation worsens in the vicinity of the discontinuity of
$\psi_{\textrm{R}}$. In the vicinity of $x\simeq22.5$, the $N=2$
results are worse than the $N=1$ results; moreover, in the $N=2$
case $\psi_{\textrm{S}}$ as well as $\psi_{\textrm{R}}$ exhibits a
discontinuity.
%
\section{Summary}
\label{summary}

\noindent In this paper we have presented a formulation of complex
time-dependent WKB (CTDWKB) that allows the incorporation of
interfering contributions to the wavefunction. The central idea in
CTDWKB presented by Boiron and Lombardi\cite{boiron} is to include
both the amplitude and the phase in the  lowest order term of the
conventional time-dependent WKB method. The rationale behind this
substitution is to treat the phase and the amplitude on equal
footings in the limit $\hbar\rightarrow0$. The benefits of the
method are twofold. Firstly, CTDWKB exhibits accuracy superior to
the conventional TDWKB\cite{boiron}. Secondly, no singularities
appear in the integration of the equations of motion. The method has
two main \textit{drawbacks}. First, the trajectories that emerge
obey the classical equations of motion but propagate in the complex
plane (due to complex initial conditions), requiring analytic
continuation of the quantum wavefunction. The second drawback is
that the reconstruction of the wavefunction on the real axis
requires a root search process. This process can be eased by
exploiting the analytic mapping between initial and final position.

We have incorporated into the CTDWKB method the possibility of
contributions from multiple crossing trajectories. Boiron and
Lombardi claim (section V in reference\cite{boiron}) that they use
the root search procedure ``excluding de facto such double
contributions", although they appreciate the benefit that double
contributions have in the GGWPD formulation. As we have demonstrated
here, considering double contributions allows description of
interference effects that are missing in the Boiron-Lombardi
formulation of CTDWKB. Moreover, we have showed how to derive higher
orders terms of the approximation in a straightforward manner. This
process was applied for the derivation of a second order term in the
CTDWKB approximation. The results for $N=2$ were better than for
$N=1$ except for a small interval in the vicinity of the classical
turning point. It was also observed that even though there are no
singularities in the integration of the CTDWKB equations of motion,
a singularity appears in the real branch $\psi_{\textrm{R}}$ at
intermediate times. For $N=2$ an irregularity also appears in the
part of the wavefunction associated with the secondary branch
$\psi_{\textrm{S}}$. We demonstrated that when a singularity appears
in $\psi_{\textrm{R}}$ (at intermediate times), the real trajectory
acts as the boundary between two groups of trajectories associated
with the real branch. Each of these groups contributes to a
different side of the singularity.

The CTDWKB formulation has several issues that require more
comprehensive study. The most critical issue is to give an analytic
explanation of the need to include the contributions from multiple
classical trajectories (with zero relative phase) and why in some
cases these contributions diverge. This will be dealt with in our
forthcoming publication \cite{newgst}. Some insight into the
analytic structure of the complex classical trajectories was given
in reference \cite{huber2} in the context of GGWPD; however, we
believe that a more general understanding of this structure is yet
to be developed. This structure presumably is relevant to the
question of when the CTDWKB  formulation converges to the exact
quantum mechanical result. We saw that in most parts of
configuration space $N=2$ performed better than $N=1$, but in other
parts of configuration space, where there were singularities, $N=2$
performed worse. What determines the position and time-dependence of
these singularities in $\psi_{\textrm{R}}$ at intermediate times?
What is the relation between the singularities in CTDWKB vs.
conventional time-dependent WKB? Is there any fundamental limitation
on the time scale for which the method is accurate? Since WKB plays
such a central role in quantum mechanics in general and in
semiclassical mechanics in particular, we believe that these
questions are of great general interest. The developments described
in this paper together with the answers to some of the above
questions could make the time-dependent WKB formulation a
competitive alternative to current time-dependent semiclassical
methods.

We wish to acknowledge David Kessler and Uzi Smilansky for useful
discussions. This work was supported by the Israel Science
Foundation $(576/04)$.

%

\end{document}